\documentstyle[seceq,mbf,wrapft]{ptptex}

\pubinfo{Vol.~97, No.~1, January 1997}  

\title{
The Minimum Mass of the First Stars 
and the Anthropic Principle
}

\author{
Takashi {\sc Nakamura}$^{1}$\footnote{
E-mail address: takashi@yukawa.kyoto-u.ac.jp
},  
 Hideya {\sc Uehara}$^{2}$\footnote{
E-mail address: uehara@tap.scphys.kyoto-u.ac.jp}
and Takeshi {\sc Chiba}$^{1}$\footnote{
E-mail address: chiba@yukawa.kyoto-u.ac.jp}
}

\inst{
$^{1}$Yukawa Institute for Theoretical Physics, 
Kyoto University, Kyoto 606-01
\\
$^{2}$Department of Physics, 
Kyoto University, Kyoto 606-01
}

\abst{
The  lower limit of the mass of the first stars suggested recently may
imply the formation of massive stars of mass greater than 8 $M_\odot$
 irrespective of the details of the initial mass function. 
The production of heavy metals from the
first stars will provide a requisite for  the existence of life without
the anthropic principle.
}

\begin{document}

\maketitle

\begin{center}
{(to be published in  {\it Progress of Theoretical Physics} 1997
January issue)
}\end{center}

\makeatletter
\if 0\@prtstyle
\def\asp{.3em} \def\bsp{.26em}
\else
\def\asp{.3em} \def\bsp{.3em}
\fi \makeatother

\section{Introduction}

It was  recently shown that  the initial mass function (IMF)
in the Galactic disk  increases monotonically with decreasing mass
and is reasonably well described by a power law $n(m) \propto
 m^{-\alpha_{disk}}$ with $\alpha_{disk} \sim 2\pm 0.5$ from 
$\sim 0.6 M_\odot$ 
down to the hydrogen-burning limit\cite{imf}.  The bimodal IMF with a
turn-over around $\sim 0.2 M_\odot$\cite{scalo} is shown to be due to 
unresolved
 binaries in the photometric luminosity function.

In the Galactic disk, grains made from heavy metals are 
the main coolants in the star formation process.
While in the formation of the first stars after the 
recombination of the universe,
hydrogen molecules are the only coolants for temperature below 
2000K since in the Big Bang Nucleosynthesis no nuclei heavier 
than B$^8$ are formed.   This implies that the star formation process of 
the first stars may be different from the present one.
 For example, the power index of the IMF of the first stars $\alpha_{first}$
 may be  completely different from $\alpha_{disk}$. 
 
The lower end of the mass of the stars 
in the present IMF is not known either. 
One might think that the lower end should be $\sim 0.08
M_\odot$ because this is the lower limit of the hydrogen-burning star.
However   in the Hayashi phase,  the pre-main sequence star itself
does not know if the hydrogen is ignited in the end. 
At present there is no reason to believe 
that only stars of mass larger than $\sim 0.08M_\odot$ are formed.
In reality the existence of  brown dwarfs (=low mass stars of 
mass less than $\sim 0.08M_\odot$ ) 
has been confirmed observationally\cite{nakaj} and
microlensing events opened the possibility that the halo of our 
Galaxy might consist of  such dim brown dwarfs\cite{macho}.

\section{The minimum mass of the first  stars and the anthropic principle}

As for the first stars,
we  know neither the IMF nor the lower end of the mass.
 Then what guarantees the formation 
of the massive  first stars of  mass $M > 8 M_\odot$  
which can scatter  metals  into the  interstellar gas ?
 One might wonder  why this is a problem.
 However the following hypothetical universe
will make the problem clear.
 Suppose that if the IMF of the first stars is a steep power law with
$\alpha_{first}$ greater than 5 and the lower end of the mass  $m_l$ is 
$0.01M_\odot$, which is suggested  through  the
analysis of the spherically symmetric collapse of the primordial 
gas cloud\cite{pss}.
Then the number of stars of mass heavier than $\sim 8M_\odot$ is given 
by
\begin{equation}
N(>8M_\odot)=C\int_8^{m_u} m^{-\alpha_{first}}dm,
\end{equation}
where $m_u$ is the upper mass limit. 
The numerical constant $C$ is determined by the total mass of a 
galaxy
\begin{equation}
M_{tot}=C\int_{m_l}^{m_u}m^{-\alpha_{first}}mdm.
\label{mass}
\end{equation}
Since we consider the index of $\alpha_{first}> 2$, the integral
 in Eq.(\ref{mass}) is insensitive to the upper end of the mass.
If we consider a typical galaxy 
of mass $M_{tot}\sim 10^{11}M_\odot$, then we have
\begin{equation}
N(>8M_\odot)  \simeq {\alpha_{first}-2\over \alpha_{first}-1}
M_{tot}m_l^{\alpha_{first}-2}8^{-(\alpha_{first}-1)}< 20
\end{equation} 
for $\alpha_{first} > 5$ and $m_l \sim 0.01M_\odot$. 
This suggests that few supernovae
occur and only a small amount of  
heavy elements are returned to the interstellar gas.
The final destiny of this hypothetical universe is the assembly of 
cooled brown dwarfs, white dwarfs and the diffuse hydrogen and helium
 gas.
No habitable planets, and therefore, no human beings exist 
in this universe. The anthropic 
principle may reject  this  hypothetical universe\cite{anth}. 
In the universe with such
a steep power index of the IMF, no human being would exist. Thus 
 in the actual universe  $\alpha_{first}$ 
must be sufficiently small.

 Recently analyzing the fragmentation process in the primordial gas cloud,
 Uehara et al.\cite{apjl} suggest that the minimum
mass of the first stars is essentially given by the Chandrasekhar mass 
$\alpha_G^{-3/2}m_p \sim 1M_\odot$, where  $m_p$ is the proton mass 
and $\alpha_G\equiv Gm_p^2/hc$.
The argument of Uehara et al. is as follows.
The primordial gas cloud collapses losing its thermal energy by 
line emissions due to hydrogen molecules.  When the collapse
proceeds enough, the collapsing timescale $t_{dyn}$ becomes equal to 
the cooling timescale $t_{cool}$.
For the cylindrical cloud with  line density $M$ and scale
radius $R$, $t_{cool}$ is estimated by
\begin{equation}
t_{cool} \sim  \frac{ \frac{1}{\gamma-1} \frac{M}{\mu m_{\rm H}} k_B T}
              {2\pi R \sigma T^4 \frac{\Delta \nu}{\nu} \alpha_c } ,
\label{A1}
\end{equation}
where $\mu, \sigma and \alpha_c$ are the mean molecular weight, the
Stefan-Boltzmann constant, and the effective number of
line emissions, respectively. The cloud temperature and 
the line profile are estimated as
\begin{equation}
  k_B T=\frac{1}{2}\mu m_{\rm H} G M, \;\;\;
\frac{\Delta \nu}{\nu}=\frac{v_{\rm H_2} }{c}
=\sqrt{ \frac{k_B T}{m_{\rm H} c^2} }.
\end{equation}
The collapsing cloud fragments when the collapse is almost halted 
\cite{IM92}, i.e., when the condition
\begin{equation}
t_{dyn} \sim t_{frag}     \label{s1}
\end{equation}
is satisfied, where $t_{frag} = 2.1 /  \sqrt{2\pi G \rho_0}$ 
is the time scale of fragmentation \cite{N87}.
From Eqs.(2.4) to (2.6)  we obtain the minimum mass of the fragment as
\begin{eqnarray}
M_{frag}  \sim  2\pi R M    
\sim  \alpha_G^{-3/2}m_p  \;.
\end{eqnarray}
The difference between the results of Uehara et al.\cite{apjl} and
those of Palla et
al.\cite{pss}
 is that 
the former estimates the mass when the condition (\ref{s1}) is satisfied,
while the latter does so when the Jeans mass becomes a minimum.

If the lower end of the mass of the first stars 
is $\sim 1M_\odot$ as suggested  by Uehara et al.\cite{apjl},  
even for the steep  IMF 
with  $\alpha_{first}=$ 5, the number of stars of 
mass heavier than $\sim 8M_\odot$ is 
\begin{equation}
N(>8M_\odot)\simeq{3\over 4}10^{11}8^{-4} > 10^7
\end{equation}
for the typical galaxy 
of mass $\sim 10^{11}M_\odot$, which is large enough to make the metal 
abundance Z  $\sim 10^{-4}$. 
Therefore the  above hypothetical universe is rejected without the 
anthropic principle, that is,  even if we do not know the details of
 the IMF of the first stars\footnote{Actually, $\alpha_{first}$ has
only to satisfy $\alpha_{first} < 13 +
\log\bigl({M_{total}\over 10^{11}}\bigr)$ in order to ensure  the existence
of massive stars.}, 
 it is certain that 
the interstellar gas is polluted by the metals, 
 which is a requisite for the existence of life.

\vskip 0.5cm

This work was supported in part by the Japanese Grant-in-Aid for Scientific
 Research Fund, No.6894 and  by
the Japanese Grant-in-Aid of Scientific Research of the Ministry of Education,
Culture, Science and Sports, No.07640399.

\end{document}